\documentclass[twocolumn]{aastex62}

\newcommand{\presto}{{\texttt{PRESTO}}}

\usepackage{amsmath}

\usepackage{longtable}

\usepackage{lineno}
\usepackage{float}

\graphicspath{{./}{figures/}}

\shorttitle{A Statistical Analysis of Crab Pulsar Giant Pulse Rates}
\shortauthors{Doskoch et al.}

\begin{document}

\title{A Statistical Analysis of Crab Pulsar Giant Pulse Rates}

\correspondingauthor{Graham\,M.\,Doskoch}
\email{gd00010@mix.wvu.edu}

\author[0000-0002-4219-6908]{Graham\,M.\,Doskoch}
\affiliation{Department of Physics and Astronomy, West Virginia University, Morgantown, WV 26506, USA}
\affiliation{Center for Gravitational Waves and Cosmology, West Virginia University, Chestnut Ridge Research Building, Morgantown, WV, 26505, USA}

\author[0000-0001-8107-8325]{Andrea\,Basuroski}
\affiliation{Department of Mechanical and Aerospace Engineering, North Carolina State University, Raleigh, NC, 27606, USA}

\author[0000-0001-8432-5889]{Kriisa\,Halley}
\affiliation{Department of Physics and Astronomy, West Virginia University, Morgantown, WV 26506, USA}
\affiliation{Center for Gravitational Waves and Cosmology, West Virginia University, Chestnut Ridge Research Building, Morgantown, WV, 26505, USA}

\author[0000-0001-7249-9822]{Avinash\,Sookram}
\affiliation{Department of Physics, The Pennsylvania State University, Abington Campus, Abington, PA, 19001, USA}

\author[0000-0002-3764-1798]{Iliomar\,Rodriguez-Ramos}
\affiliation{Department of Physics, University of Puerto Rico at Mayagüez, Mayagüez, Puerto Rico, 00681, USA}

\author[0000-0003-2162-7286]{Valmik\,Nahata}
\affiliation{New Providence High School, New Providence, NJ, 07974, USA}

\author{Zahi\,Rahman}
\affiliation{Sanford H. Calhoun High School, Merrick, NY, 11566, USA}

\author[0000-0003-3614-5458]{Maureen\,Zhang}
\affiliation{Pittsford Sutherland High School, Pittsford, NY, 14534, USA}

\author{Ashish\,Uhlmann}
\affiliation{Newton South High School, Newton, MA, 02459, USA}

\author[0000-0001-6840-5117]{Abby\,Lynch}
\affiliation{Sachem North High School, Lake Ronkonkoma, NY, 11779, USA}

\author[0000-0003-0771-6581]{Natalia\,Lewandowska}
\affiliation{Department of Physics, State University of New York Oswego, Oswego, NY, 13126, USA}

\author{Nohely\,Miranda}
\affiliation{U.S. Navy, 6149 Welsh Rd, Dahlgren, VA, 22448, USA}

\author{Ann\,Schmiedekamp}
\affiliation{Department of Physics, The Pennsylvania State University, Abington Campus, Abington, PA, 19001, USA}

\author{Carl\,Schmiedekamp}
\affiliation{Department of Physics, The Pennsylvania State University, Abington Campus, Abington, PA, 19001, USA}

\author{Maura\,A.\,McLaughlin}
\affiliation{Department of Physics and Astronomy, West Virginia University, Morgantown, WV 26506, USA}
\affiliation{Center for Gravitational Waves and Cosmology, West Virginia University, Chestnut Ridge Research Building, Morgantown, WV, 26505, USA}

\author{Daniel\,E.\, Reichart}
\affiliation{Skynet Robotic Telescope Network,University of North Carolina, Chapel Hill, NC 27599, USA}

\author{Joshua\,B.\, Haislip}
\affiliation{Skynet Robotic Telescope Network,University of North Carolina, Chapel Hill, NC 27599, USA}

\author{Vladimir\,V.\, Kouprianov}
\affiliation{Skynet Robotic Telescope Network,University of North Carolina, Chapel Hill, NC 27599, USA}

\author{Steve\, White}
\affiliation{Green Bank Observatory, Green Bank, WV 24944, USA}

\author{Frank\, Ghigo}
\affiliation{Green Bank Observatory, Green Bank, WV 24944, USA}

\author{Sue\, Ann\, Heatherly}
\affiliation{Green Bank Observatory, Green Bank, WV 24944, USA}

\begin{abstract}
A small number of pulsars are known to emit giant pulses, single pulses much brighter than average. Among these is PSR J0534+2200, also known as the Crab pulsar, a young pulsar with high giant pulse rates. Long-term monitoring of the Crab pulsar presents an excellent opportunity to perform statistical studies of its giant pulses and the processes affecting them, potentially providing insight into the behavior of other neutron stars that emit bright single pulses. Here, we present an analysis of a set of 24,985 Crab giant pulses obtained from 88 hours of daily observations at a center frequency of 1.55~GHz by the 20-meter telescope at the Green Bank Observatory, spread over 461 days. We study the effects of refractive scintillation at higher frequencies than previous studies and compare methods of correcting for this effect. We also search for deterministic patterns seen in other single-pulse sources, possible periodicities seen in several rotating radio transients and fast radio bursts, and  clustering of giant pulses like that seen in the repeating fast radio burst FRB121102.
\end{abstract}


\section{Introduction} \label{sec:intro}

The Crab pulsar has been of significant interest to astronomers throughout the half century since its discovery \citep{Staelin+Riefenstein1968}. As a young, bright, radio-loud neutron star still embedded in its supernova remnant and pulsar wind nebula, it has provided numerous opportunities to understand the complex dynamics of the system across the electromagnetic spectrum (for reviews, see \citealt{Hester2008} and \citealt{Buhler+Blandford2014}). Additionally, the Crab is one of the few pulsars known to emit radio giant pulses (GPs), sporadic single pulses several orders of magnitude stronger than its regular emission \citep{Hankins+2003,Cordes+2004}.

Decades of study have revealed much about the Crab pulsar's giant pulses: they are broadband \citep{Sallmen+1999}, their amplitudes follow a power-law distribution (see e.g. \citealt{Argyle+Gower1972,Popov+Stappers2007,Mickaliger+2012}), and some may have intrinsic widths on the order of a nanosecond \citep{Hankins+2003}, with brightness temperatures reaching $\sim10^{41}$ K \citep{Hankins+Eilek2007}. They are known to be emitted at the main pulse or interpulse (see e.g. \citealt{Lundgren1994}) with additional giant pulses observed at two high-frequency components up to $\sim$8.4~GHz \citep{Jessner+2005}. However, giant pulses are not detected every rotation period, and therefore do not immediately appear periodic. Over long timescales, Crab giant pulses appear to behave like a Poisson process \citep{Lundgren+1995,Karuppusamy+2010}. Observed mean pulse rates vary depending on telescope sensitivity and observing frequency \citep{Rankin+1970} and may be anywhere from several to tens of pulses per minute at L-band and higher frequencies \citep{Karuppusamy+2010,Mickaliger+2012} to hundreds of pulses per minute at 430 MHz \citep{Cordes+2004}.

Over the past two decades, two new classes of single-pulse-emitting radio sources have been discovered, garnering significant attention. One is rotating radio transients (RRATs; \citealt{McLaughlin+2006}), pulsars detected only through sporadic single pulses, rather than through their time-averaged, periodic emission. Myriad models for the intermittency of this emission have been proposed, including fallback of material from supernova remnants \citep{Li2006}, radiation belts \citep{Luo+Melrose2007}, pulsars undergoing extreme nulling \citep{Wang+2007,BurkeSpolaor2013}, and circumstellar material \citep{Cordes+Shannon2008}; however, the reason RRAT activity is so sporadic remains unknown. The other class of objects is fast radio busts (FRBs; \citealt{Lorimer+2007}), millisecond-duration bursts of extragalactic origin. Many FRBs appear to be one-off isolated events, while a small fraction repeat \citep{Spitler+2016}.  The mechanisms behind FRB emissions remain unknown, although they may involve neutron stars (see e.g. \citealt{Platts+2019} for a full list of theories\footnote{\url{https://frbtheorycat.org/}}). A popular set of models involves bursting magnetars, akin to the radio burst observed from the Galactic magnetar SGR J1935+2154 in 2020, although it is difficult for this model to explain all FRBs \citep{CHIMEMagnetar}.

Similarities between FRBs and Crab giant pulses have inspired the proposal that FRBs may be giant or supergiant pulses from young, extragalactic neutron stars \citep{Cordes+Wasserman2016,Connor+2016}. Subsequent studies have found additional common characteristics, including their energy distributions \citep{Lyu+2021} and narrow-band features \citep{Thulasiram+Lin2021}, and encouraged the the development of unified emission models (see e.g. \citealt{Lyutikov2021}). This motivates characterization of any properties of Crab giant pulses akin to those seen in FRBs and other single-pulse sources.

One important property for comparison is the burst or pulse emission rates and how they vary with time. Several phenomena appear in the pulse rates or burst activity of RRATs and FRBs. The first is a stochastic modulation in time due to interstellar scintillation \citep{Spitler+2018}. The Crab pulsar is also known to experience a modulation in its observed flux density on several frequency-dependent timescales due to refractive and diffractive scintillation in the interstellar medium (see e.g. \citealt{Rickett+Lyne1990} and \citealt{Cordes+2004}). It is expected that individual Crab giant pulses should be similarly modulated, inducing the stochastic fluctuations observed in the Crab's giant pulse rate.

Second, some single-pulse sources exhibit periodicity in burst rates. There is tentative evidence that the pulse rates of the RRATs J1819--1458 and J1754--3014 vary on timescales of $\mathcal{O}(1000)$ days \citep{Palliyaguru+2011}. More recently, it has been shown that FRB180916 follows a cycle of activity lasting 16.35 days \citep{CHIME2020}, and a periodicity of $\sim160$ days has been found in the burst rate of the highly active repeater FRB121102 \citep{Rajwade+2020,Cruces+2021}. This is of interest given that possible periodicities of 0.41 and 0.99 days have been seen in the Crab's giant pulse rate \citep{Mickaliger+2012}, but have not been confirmed; the different timescales imply different processes may be at work.

Finally, the times between bursts for some FRBs are decidedly non-Poissonian. For example, the wait times of FRB121102 are better described by a Weibull distribution \citep{Opperman+2018}, featuring clumping in burst arrival times, which could be due to scintillation \citep{Spitler+2018}. Crab giant pulses are typically modeled as a Poisson process \citep{Lundgren+1995,Karuppusamy+2010}, but this assumption is worth testing.

In this paper, we present the results of a daily observing campaign of the Crab pulsar at a center frequency of 1.55~GHz, with the goal of searching for similar patterns in giant pulse rates. Daily observations spread over 461 days yielded a total of 24,985 giant pulses, which we used to search for stochastic and deterministic changes in giant pulse rates. In Sections~\ref{sec:observations} and \ref{sec:pipeline} we discuss our observations and the single-pulse pipeline used to extract giant pulses from them. In Section~\ref{sec:riss} we present a short theoretical overview of refractive scintillation and its effects on giant pulse observations, and in Section~\ref{sec:corrections} we compare two methods of correcting for it. In Section~\ref{sec:structure-functions} we compute basic scintillation quantities and compare them to theory and previous work. In Section~\ref{sec:periodicity} we present a search for periodicity in the Crab's giant pulse rate, and in Section~\ref{sec:wait-times} we search for giant pulse clustering in time and relationships between pulse strength and wait time. In Section~\ref{sec:event} we discuss an excess of giant pulses observed in early December 2022. We conclude in Section~\ref{sec:conclusion}.

\begin{figure*}
    \centering
    \includegraphics[width=1.9\columnwidth]{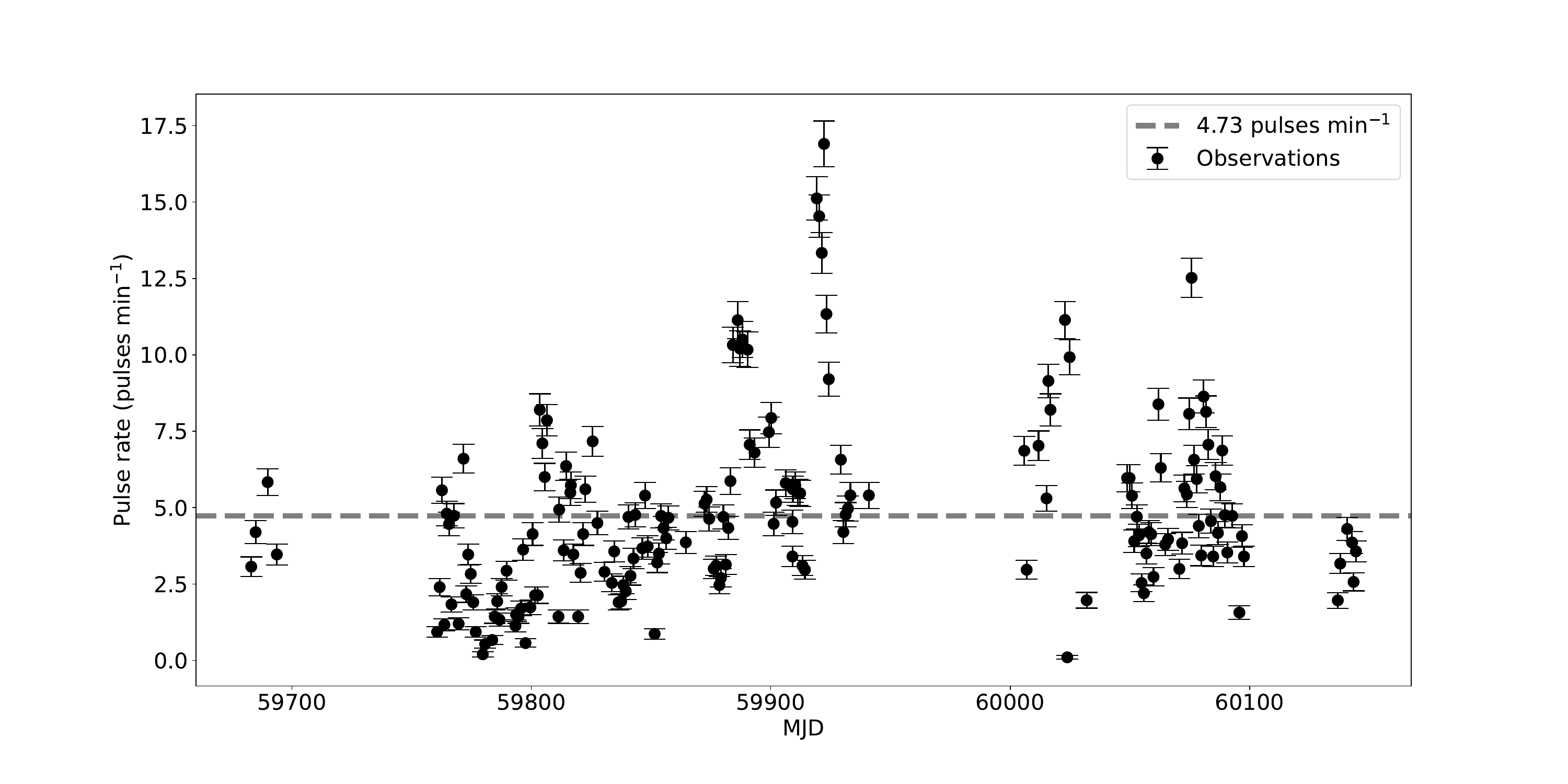}
    \caption{Crab giant pulse rates from the observing campaign presented here. All uncertainties assume Poisson statistics, i.e. the uncertainty in observing $N$ pulses is $\sqrt{N}$ pulses. The mean pulse rate is shown as a horizontal grey dashed line.}
    \label{fig:pulse-rates}
\end{figure*}

\section{Observations} \label{sec:observations}

The processes of interest take place on a variety of timescales. Candidate periodicities in some FRBs and RRATs last on the order of hundreds to thousands of days. Refractive scintillation, on the other hand, can act on timescales as short as days when observing at frequencies above several GHz. Probing all of these possible phenomena necessitates a long-term, high-cadence observing campaign. We therefore performed daily observations of the Crab with the 20-meter telescope at the Green Bank Observatory. All observations used the telescope's L-band receiver, centered at 1.55~GHz, paired with the CYBORG backend, yielding 288~MHz of bandwidth divided into 1024 frequency channels, with a time resolution of 1.04 ms. This time resolution is not optimal for this study, but we were constrained by limitations of the CYBORG backend and data storage. The data were stored as 16-bit FITS files.

Observations began on April 13, 2022, with the final observation in this data set taken on July 19, 2023. No observations were taken in May of 2022 while we tested our pipeline on the first $\sim$10 days of observations. Few observations were taken during June of either year to avoid any effects from the Crab passing close to the Sun during that time. There were additional interruptions during the campaign lasting weeks to months due to telescope maintenance, including a long maintenance window starting at the end of 2022 and continuing into March 2023.

Miscellaneous issues caused some observations to be rescheduled or cancelled, leading to multiple observations on certain days. All observations lasted $\sim30$ minutes, a length chosen because it enabled us to detect enough giant pulses to perform robust statistical analysis on individual observations while using minimal telescope time and reducing data storage requirements. Details of individual observations can be found in Table~\ref{tab:observation-details} in Appendix~\ref{sec:observation-appendix}.

\section{Analysis} \label{sec:pipeline}

\begin{figure*}
    \centering
    \includegraphics[width=1.5\columnwidth]{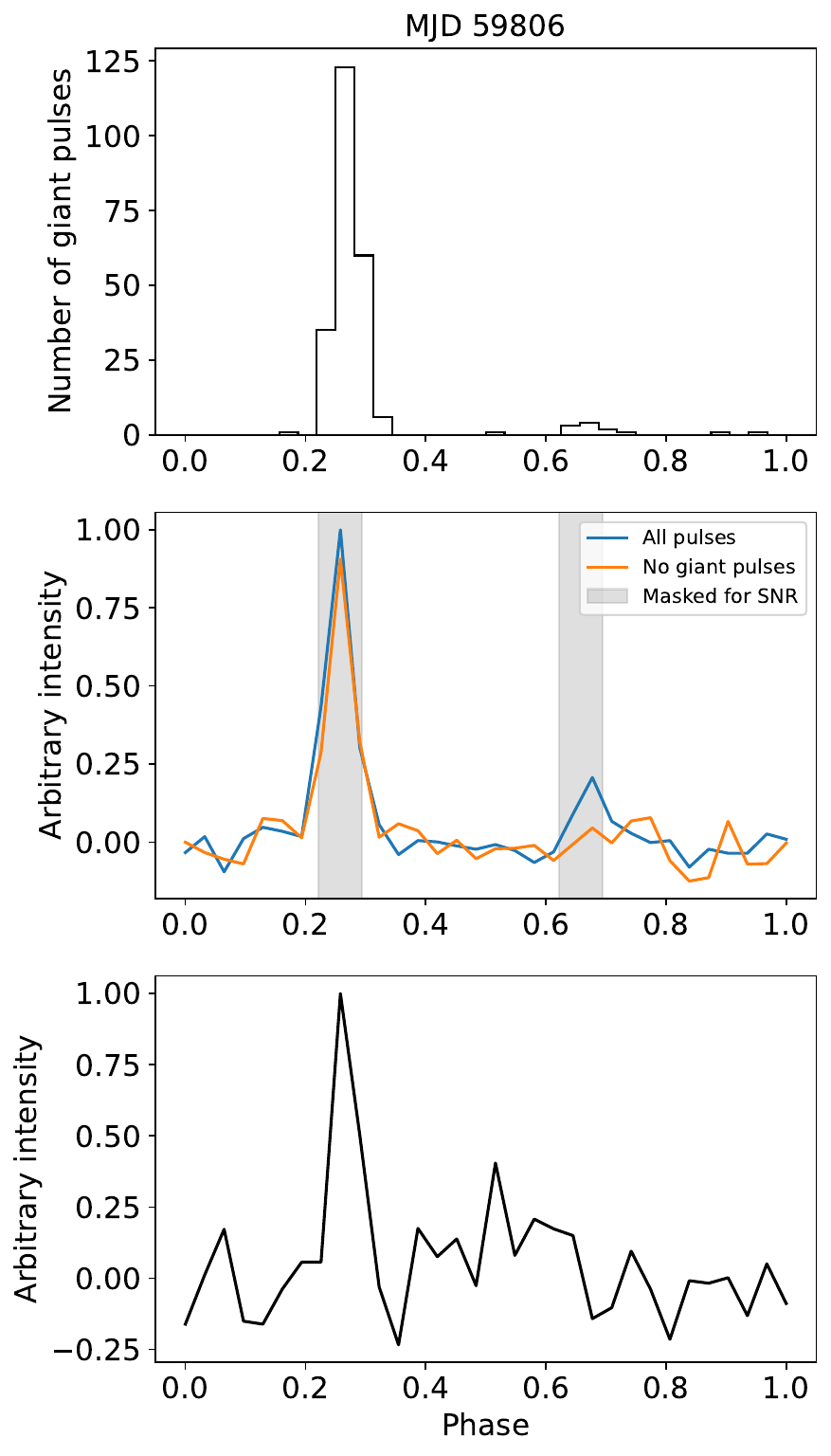}
    \caption{\textit{Top:} Histogram of phases of giant pulse candidates from the observation on MJD 59806. There are clear peaks at the main pulse and interpulse. Several candidates fall outside these regions and are likely spurious. \textit{Middle:} Folded profiles from the observation, one with giant pulses and one made with only a set of single pulses. The interpulse appears to be weaker without the giant pulses. In gray are the on-pulse regions, which include the three bins surrounding the main pulse and interpulse of the profile; the off-pulse regions were used for computing noise levels when calculating profile SNR. \textit{Bottom}: Profile of a typical giant pulse from the observation.}
    \label{fig:profiles}
\end{figure*}

The data were searched for giant pulses using a pipeline based on the \presto\footnote{\url{https://github.com/scottransom/presto}} package \citep{Ransom2011}. Radio frequency interference (RFI) was identified and removed by the \texttt{rfifind} routine. Each observation was then folded at the Crab's spin period to generate a profile using the ephemeris from the Jodrell Bank Centre for Astrophysics\footnote{\url{https://www.jb.man.ac.uk/pulsar/crab.html}} \citep{Lyne+1993}. The folded profile was assigned a signal-to-noise ratio (SNR), computed as the ratio of the peak on-pulse intensity divided by the standard deviation $\sigma$ of the off-pulse noise. We designated the on-pulse region as being all bins less than $\delta$ in phase away from the center of the main pulse and interpulse, with $\delta=0.036$ the duty cycle \citep{Karuppusamy+2010}, and excluded those bins when calculated $\sigma$. Figure~\ref{fig:profiles}, generated from the observation on MJD 59806, compares the folded profile, the phase distribution of giant pulses, and a single giant pulse. The on-pulse regions for the main pulse and interpulse are shaded in gray.

The observations were then manually examined. We excised 26 observations due to RFI contaminating much of the band or an SNR below a chosen threshold ($\mathrm{SNR}=6$); this was done to exclude observations likely to have large numbers of spurious pulse candidates from interference or have detections too weak to robustly contribute to the analysis. Ultimately, only one observation without substantial RFI yielded $\mathrm{SNR}<6$; we include it in the Appendix for completeness but did not use it for the analysis in Sections~\ref{sec:corrections}-\ref{sec:event}. This yielded a final data set of 176 observations lasting 88 hours.

Due to instrumental limitations, observations did not include flux calibration. As a proxy for the Crab's flux density on a given day, we use the folded profile SNR divided by the square root of the length of the observation $T_{\mathrm{obs}}$, since we expect $\mathrm{SNR}\propto T_{\mathrm{obs}}^{1/2}$ from the radiometer equation.\footnote{Since all observations are approximately the same length, with deviations on the order of seconds, the effects of this correction are miniscule, but we include it for completeness.} Likewise, giant pulse amplitudes are represented as SNRs instead of fluences.

The data were dedispersed at 11 different dispersion measures (DMs; first used in the context of pulsars by \citealt{Hewish+1968}) centered on the Crab pulsar's DM at each epoch, in steps of $1$ pc cm$^{-3}$, using the \presto\;routine \texttt{prepsubband}. The step size was chosen to ensure pulses are astrophysical by requiring them to peak at the Crab's DM. This central DM was taken to be the DM at each epoch from the Jodrell Bank ephemeris. Each dedispersed time series was subsequently searched for single pulses using the \presto\;routine \texttt{single\_pulse\_search.py}. In each time series, when two candidate pulses were separated by less than some time difference $\Delta$, we discarded the weaker of the two. We used $\Delta=6.75$ ms, a value we determined empirically, to avoid multiple detections of a single real giant pulses. We then kept all pulses with SNRs above 5 and whose SNRs peaked at the epoch's central DM.

\begin{figure}
    \centering
    \includegraphics[width=0.95\columnwidth]{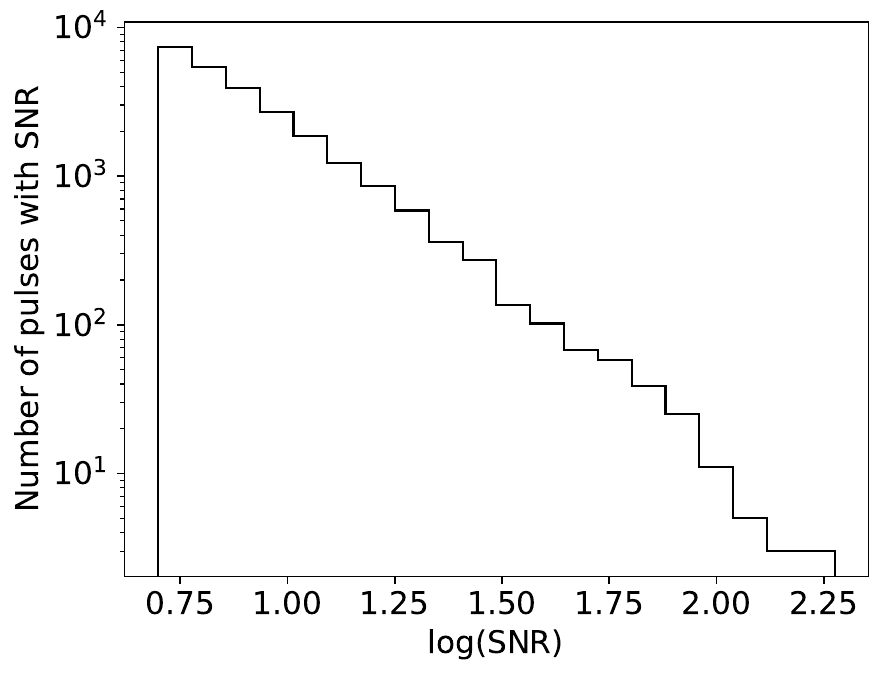}
    \caption{Histogram of giant pulse SNRs. The distribution approximately follows a power law, as expected.}
    \label{fig:snr_histogram}
\end{figure}

There is no broadly-accepted minimum SNR cutoff used for Crab giant pulses. Various values are used in the literature, including $5\sigma$ \citep{Cordes+2004}, $7\sigma$ \citep{Bilous+2011}, $8\sigma$ \citep{Main+2021} and $10\sigma$ \citep{Mickaliger+2012}. We choose $5\sigma$ because the resulting sample includes as many giant pulses as possible while, with appropriate cutoffs, rejecting spurious candidates. This is apparent upon examining the results of the pipeline from individual observations, which display power-law giant pulse SNR distributions and show pulses occurring almost exclusively at the main pulse and interpulse.

Figure~\ref{fig:pulse-rates} shows the observed giant pulse rates from the observing campaigns. The error bars in pulse rate are computed assuming Poisson statistics, in which the uncertainty in the number of pulses detected is the square root of the number of pulses; modeling the pulse rate as a Poisson process is favored by our analysis in Section~\ref{sec:wait-times}. Figure~\ref{fig:snr_histogram} shows the SNR distribution of all giant pulses in the sample. As expected, it approximately follows a power law, with apparent deviations at high SNRs potentially due to small-number statistics. Figure~\ref{fig:profiles} shows the distribution of giant pulses in a single observation, as well as folded profiles of the Crab and a profile of a typical giant pulse.

% GD: The profile without giant pulses doesn't include the entire observation, just most of it. I added together the profiles of the different segments by aligning them based on which bin had the highest value in each profile. For most short intervals, though, there was noise, and so that technique did not correspond to the actual peak of the pulse. This led to an artificially strong profile. I've only included intervals here of at least ~10 seconds, which still encompasses most of the observation and ensures I'm properly adding the profiles together.

\section{Refractive scintillation} \label{sec:riss}

The interstellar medium (ISM) is turbulent and highly inhomogeneous. Scattering by these inhomogeneities causes scintillation \citep{Rickett1990,Narayan1992}, a phenomenon that manifests in observations of pulsars as stochastic modulations of the observed pulse strength \citep{Sieber1982}. Small-scale inhomogeneities induce diffractive interstellar scintillation (DISS), with intensity variations on timescales of minutes to hours, whereas large-scale inhomogeneities induce refractive interstellar scintillation (RISS), causing pulse intensity variations on timescales of weeks to months.

There is a strong frequency dependence to these timescales. The thin-screen model \citep{Scheuer1968} predicts that the diffractive timescale should scale as $\tau_{\mathrm{DISS}}\propto f^{1.2}$ and the refractive timescale should scale as $\tau_{\mathrm{RISS}}\propto f^{-2.2}$ \citep{Lorimer+Kramer2012}. Multifrequency observations by \cite{Rickett+Lyne1990} confirmed that the refractive frequency dependence is consistent with the $f^{-2.2}$ scaling predicted by theory. At high frequencies, refractive scintillation should induce significant variation between observations in high-cadence observing campaigns. No prior study has produced robust direct estimates of the refractive timescale at GHz frequencies or above; timescales at these frequencies can be extrapolated from lower-frequency measurements or from measurements of DISS, but this requires assuming that theoretical models of turbulence are correct.

Refractive scintillation consequently induces stochastic variations in the Crab's observed giant pulse rate. If the minimum signal-to-noise threshold for a giant pulse is held constant across observations, days with the intensity modulated upwards will show higher giant pulse rates, while days with the intensity modulated downwards will show lower giant pulse rates. While this may obscure intrinsic giant pulse rate variations, it provides an opportunity to use giant pulses to probe the ISM.

\section{Correcting for scintillation} \label{sec:corrections}

It is possible that the stochastic variations in giant pulse rate due to scintillation can obscure deterministic changes, such as long-term periodicity or short-term spurts of activity. We therefore explore two methods used in the literature to correct for the stochastic modulations in giant pulse rate.

The first method, described by \cite{Mickaliger+2012}, relies on the fact that the strength of the Crab's folded profile should be modulated by the same factor and on the same timescale as individual giant pulses. For each observation, we folded the data using \presto's \texttt{prepfold} routine as described in Section~\ref{sec:pipeline}. Due to a lack of calibration scans, the SNR of the resulting profile was used as a proxy for intensity. A scaling factor for the observation was then obtained by dividing the highest profile SNR of the observing campaign (excluding observations during the event detailed in Section~\ref{sec:event} by the observation's folded profile SNR), from MJD 60075.

The second method was developed by \cite{Bilous+2011}, motivated by observations for which an adequate folded profile could not be obtained. A reference observation was chosen with 
a large SNR and a high pulse rate.\footnote{\citealt{Bilous+2011} chose the observation with the highest pulse rate; here, for consistency, we used the same observation as in the profile SNR method, as it still had one of the highest observed pulse rates.} For each other observation, pulses were divided into a set number of logarithmically-spaced SNR bins $N_{\mathrm{bins}}$. We then defined the quantity
\begin{equation}
    \chi_k^2\equiv\frac{1}{N_{\mathrm{bins}} - 1}\sum_{I_i}\frac{[N_{\mathrm{ref}}(I_i) - N(I_i/k)]^2}{\sigma_{N(I_i/k)}^2}
\end{equation}
where $N(I_i)$ is the number of pulses per hour with SNR at least $I_i$ and
\begin{equation}
    \sigma_{N(I_i/k)}\equiv\frac{\sqrt{N(I_i/k)}}{T}
\end{equation}
with $T$ the length of the observation in hours. The scaling factor for each observation was then the $k$ that minimized its $\chi_k^2$. We chose $N_{\mathrm{bins}}=20$ and searched values of $k$ ranging from $k=0.1$ to $k=20$.

There are advantages and disadvantages to both methods. The results of the first may be incorrect for observations with weak folded pulse profiles. The results of the second may be incorrect for observations with low numbers of giant pulses or where the slope of the power law distribution of giant pulses is significantly different from the slope in the reference observation.

\begin{figure}
    \centering
    \includegraphics[width=0.95\columnwidth]{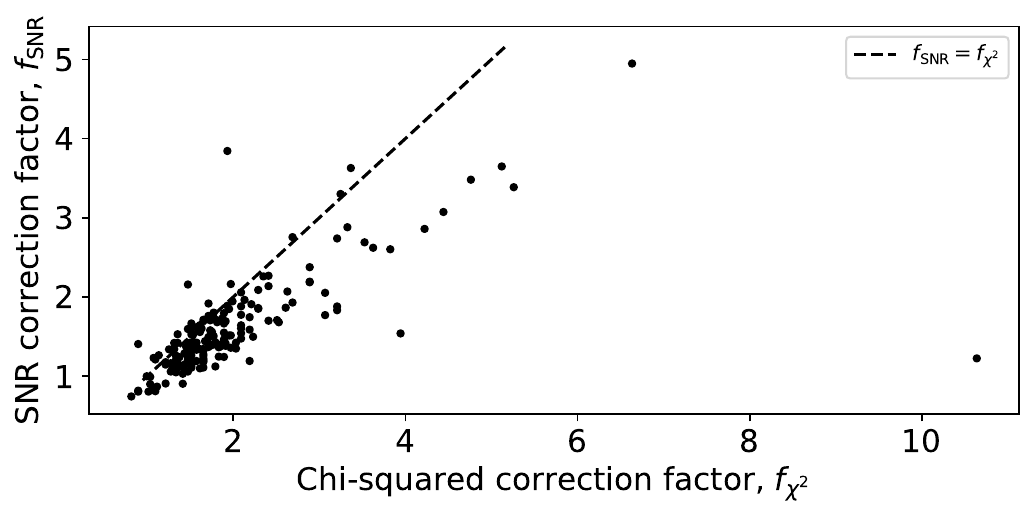}
    \caption{Comparison between the two methods of accounting for scintillation. There is a strong correlation between the sets of correction factors, but the $\chi_k^2$ method consistently yields higher results.}
    \label{fig:factor-comparison}
\end{figure}

The results of applying these two methods to the data set are shown in Figure~\ref{fig:factor-comparison}. The reference high-SNR observation is from MJD 60075. There is a strong positive correlation between both methods, although the $\chi_k^2$ method produces consistently higher corrections. While the precise source of the bias is unclear, the $\chi_k^2$ factors certainly suffer from intrinsic limitations from the observations. Most observations have no more than $\mathcal{O}(100)$ pulses, ensuring that small-number statistics affect many of the bins. This may have led to slightly poorer fits; only 124 of the 175 non-reference observations yielded $\chi_k^2\lesssim2$, although this compares favorably to the results of \citealt{Bilous+2011}.

\begin{figure}
    \includegraphics[width=0.95\columnwidth]{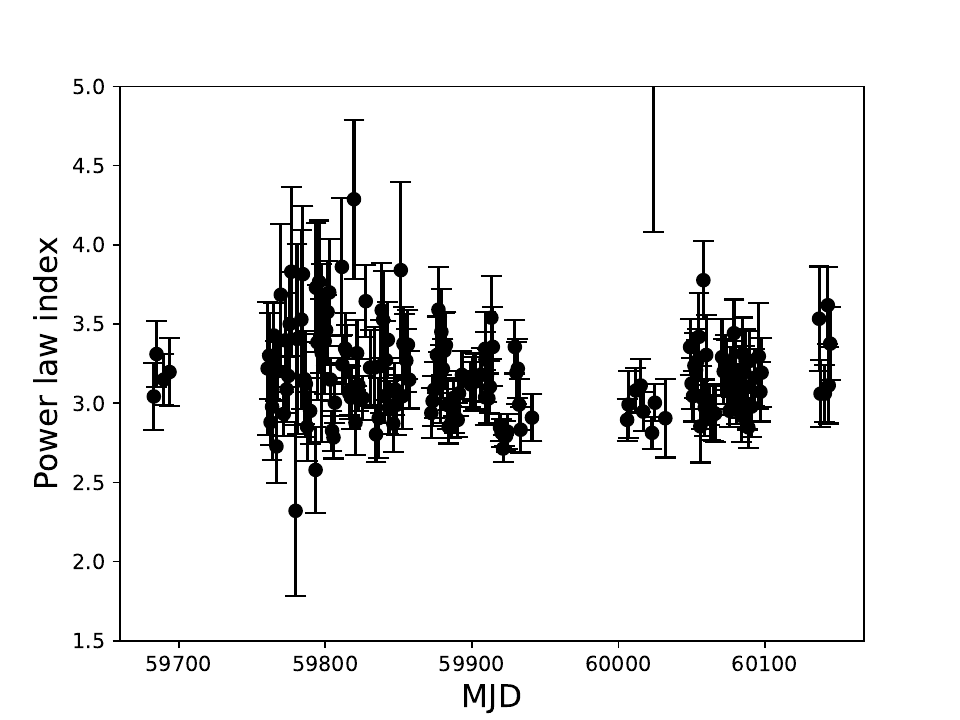}
    \caption{Time series of the power-law index of the giant pulse SNR distribution. The y-axis cuts off one observation with a poorly-constrained index of $\hat{\alpha}\simeq8$, based on few giant pulses.}
    \label{fig:power-laws}
\end{figure}

One of the key assumptions of the $\chi_k^2$ method is that the shape of the distribution of giant pulse intensity does not change. To test this, we performed a power-law fit to the giant pulses from each observation, assuming the distribution scales like $N(s)\propto \sigma^{-\alpha}$, with $N(s)$ the number of pulses with SNR $s$. We used a maximum likelihood estimator method (see \citealt{Clauset+2009} for a discussion of power-law fitting and \citealt{Oronsaye+2015} for an application to Crab giant pulses). This produces an estimate
\begin{equation}
\hat{\alpha}=1+n\left[\sum_{i=1}\ln\frac{s_i}{s_{\mathrm{min}}}\right]^{-1}
\end{equation}
with error
\begin{equation}
\hat{\sigma}=\frac{\hat{\alpha} - 1}{\sqrt{n}}
\end{equation}
where the sum is over the $n$ pulses from an observation. We chose $s_{\mathrm{min}}=5$. Figure~\ref{fig:power-laws} shows estimates of $\hat{\alpha}$ for each observation.

\begin{figure}
    \centering
    \includegraphics[width=0.95\columnwidth]{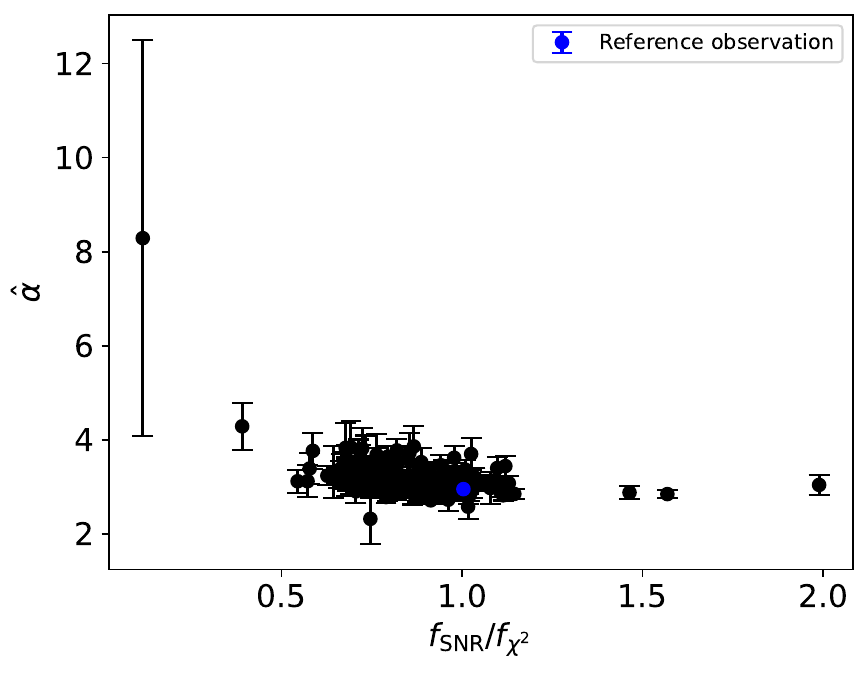}
    \caption{The relationship between the ratio of the two correction factors and the power-law index of each observation. There is a mild correlation, and changes in the shape of the giant pulse distribution are able to explain observations that showed extreme disagreement between the two methods.}
    \label{fig:power-law-variations}
\end{figure}

We find that the observations with the greatest difference between the two methods indeed tend to have unusual power-law indices, with some being much steeper than the reference observation's. As seen in Figure~\ref{fig:power-law-variations}, there is a clear correlation between power law index and the ratio of the two correct factors, though it does not explain all of the disagreement.

We do not apply the results of these methods to any analysis in this paper. First, the disagreement between the two is non-zero, which may be attributed to the observation quality reasons mentioned above. Second, the $\chi_k^2$ method always absorbs any intrinsic changes in the intensity or number of individual giant pulses. The SNR method does the same if the folded profile is dominated by giant pulses, although, as with \citealt{Mickaliger+2012}, this is not the case in our data set. This is unsurprising, as the distribution of giant pulse SNRs follows a clear power law with no turnover at low SNR. Since the refractive scintillation timescale at GHz frequencies is expected to be on the order of a couple of days, this somewhat restricts our ability to draw definitive conclusions about individual observations, but should not affect our results about trends on timescales much longer than the scintillation timescale.

\section{Structure functions and refractive timescales} \label{sec:structure-functions}

The most common tool used in the literature to study refractive scintillation is the first-order structure function. Say a stochastic process $X$ is observed over a time $T$, divided into $N$ time intervals of length $\Delta t = T/N$. The structure function of $X$ is defined for a time lag $k$ by \citep{Simonetti+1985,Stinebring+2000}
\begin{equation}
D_X(k)\equiv\frac{1}{\langle X\rangle^2N(k)}\sum_{i=1}^Nw(i)w(i+k)[X(i) - X(i + k)]^2
\end{equation}
where $X(i)$ is the mean value of $X$ observed in interval $i$, $w(i)$ is the number of observations in interval $i$, and
\begin{equation}
N(k)\equiv\sum_{i=1}^Nw(i)w(i+k).
\end{equation}
We have normalized the structure function using the square of the mean value of $X$ over $T$, $\langle X\rangle^2$, as per the convention of \cite{Stinebring+2000}.\footnote{Prior work set $w(i)=1$ if an observation was performed during interval $i$ and $0$ otherwise; due to the high cadence of our observations, we generalize the definition to include intervals with multiple observations.}

As $k\to\infty$, the structure function becomes saturated, flattening at the value $2\sigma^2$, with $\sigma^2$ the variance of $X$. The characteristic timescale $\tau_X$ is then defined as the lag at which the structure function reaches half of its saturated value. This divides the timescales into three different regimes of behavior. Short lags ($k\ll\tau_X$) fall into the noise regime. Long lags ($k\gg\tau_X$) fall into the saturation regime. In the intermediate regime, where $k\sim\tau_X$, the structure function behaves like a power law in $k$.

Traditional studies of the effects of refractive scintillation on the Crab pulsar have set $X$ to be the flux density of the folded profile (see e.g. \citealt{Rickett+Lyne1990}). We do so using the folded profile SNR as a proxy for flux density. However, as refractive scintillation should simultaneously modulate the observed giant pulse rate, we can perform the same analysis with $X$ being the observed giant pulse rate.

We computed errors in the structure function using analytical approximations derived by \citealt{Rickett+2000}. Our observing campaign satisfies $T\gg\tau_X$; for the regime where $k\gg\tau_X$, we have (Eq. B8 and B9 of \citealt{Rickett+2000})
\begin{equation}
\sigma_{D_X}(k)\sim\left[\frac{4}{T}\int_0^{\infty}(D_{X,\infty} - D_X(r))^2\;\mathrm{d}r\right]^{1/2}
\end{equation}
where $D_{X,\infty}$ is the saturated value of the structure function. For the regime where $k\simeq\tau_X$, we have (Eq. B12):
\begin{equation}
\sigma_{D_X}(k)\sim 2D_{X,\infty}\sqrt{\frac{\tau_X}{T}}\frac{k^2}{k^2+\tau_X^2}.
\end{equation}

The refractive timescale $\tau_{\mathrm{RISS}}$ is now defined as the lag at which the structure function reaches half of its saturated value. The upper and lower error bars in $\tau_{\mathrm{RISS}}$ are calculated by computing the lags at which the structure function reaches values of $\frac{1}{2}(D_{X,\infty}\pm\delta D_{X,\infty})$, with the approximation for the root-mean-square error being \citep{Rickett+2000}
\begin{equation}
\frac{\delta D_{X,\infty}}{D_{X,\infty}}=\sqrt{\frac{2\tau_{\mathrm{RISS}}}{T}}.
\end{equation}

\begin{figure}
    \centering
    \includegraphics[width=0.95\columnwidth]{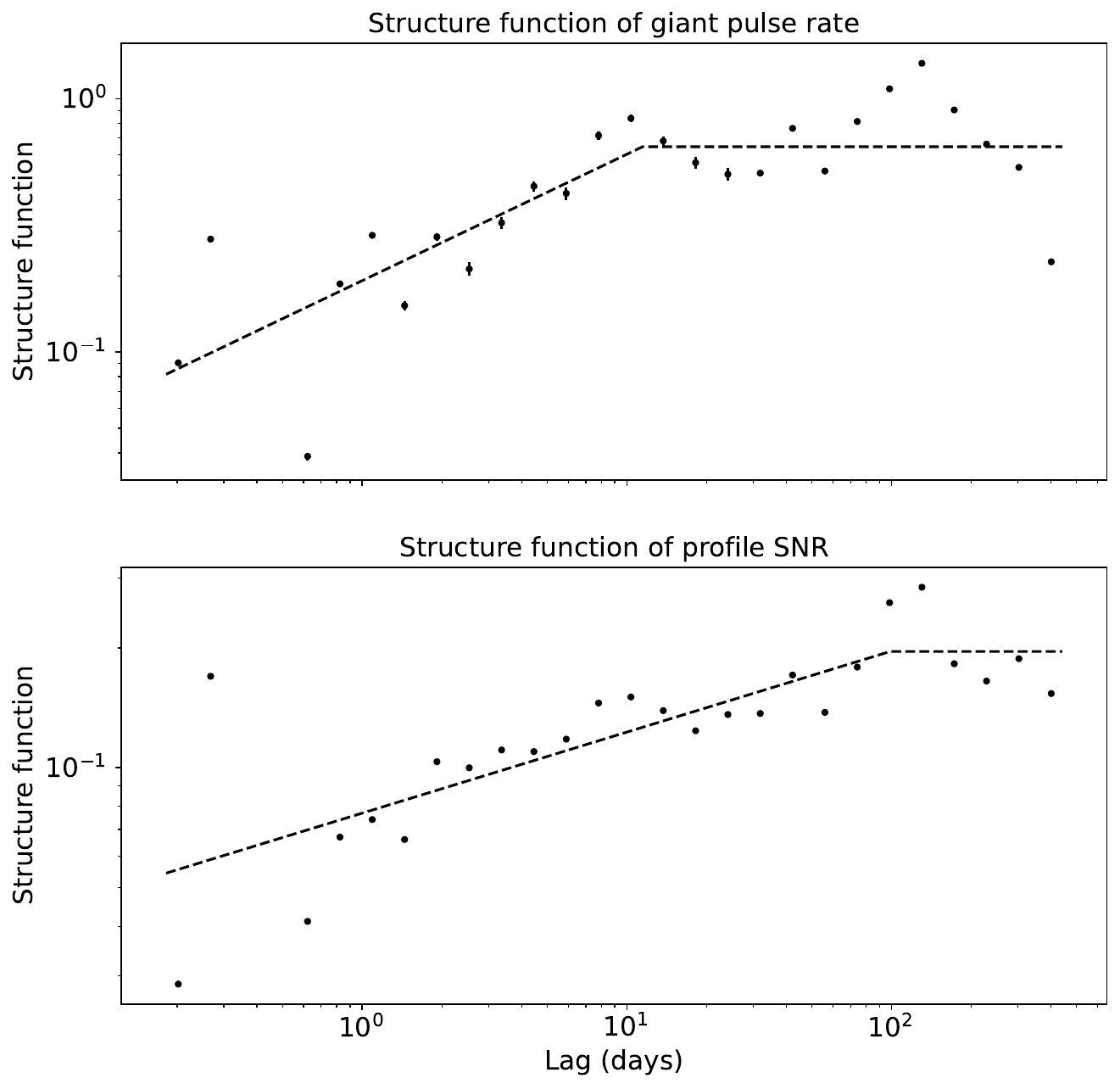}
    \caption{Structure functions for giant pulse rate and profile SNR. The dashed lines are the best-fit values of the structure function fitted as a power law followed by the saturation regime.}
    \label{fig:structure-functions}
\end{figure}

We computed the structure functions and refractive timescales for two time series: the pulse rates and folded profile SNRs (scaled by $T_{\mathrm{obs}}^{-1/2}$). As noted earlier, we excluded any observations with unscaled profile SNRs less than 6, to conservatively exclude poorer observations. Figure~\ref{fig:structure-functions} shows the structure functions and fits for both giant pulse rate and profile SNR.

\begin{figure}
    \centering
    \includegraphics[width=0.95\columnwidth]{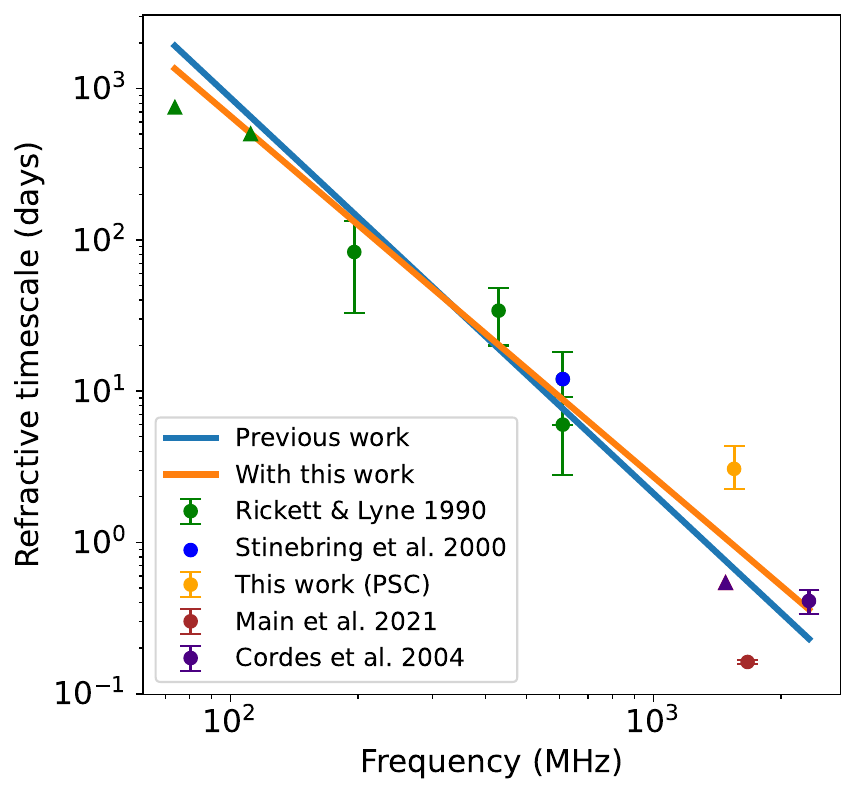}
    \caption{Frequency scaling of the refractive timescale, with power-law fits when using previous studies and then including this work. Points with error bars are measurements using the structure function method or by extrapolating from diffractive scintillation; triangles represent lower limits.}
    \label{fig:timescale-comparison}
\end{figure}

We derive timescales of $2.85_{-0.60}^{+0.67}$ and $4.27_{-1.51}^{+2.41}$ days from the time series of giant pulse rate and profile SNR, respectively, which agree within uncertainty. The timescales and uncertainties are listed in Table~\ref{tab:previous-scintillation}, along with previous measurements, direct and indirect, of $\tau_{\mathrm{RISS}}$ for the Crab pulsar across a range of frequencies. Figure~\ref{fig:timescale-comparison} shows the values from previous  studies and our values as a function of observing frequency. We fit a power-law (i.e. $\tau_{\mathrm{RISS}}\propto f^{\alpha}$) only to previous work and then to previous work with our values added, with our $\tau_{\mathrm{RISS}}$ taken to be the average of the two methods. We find $\alpha=-2.61\pm0.38$ for previous work and $\alpha=-2.38\pm0.41$ when including our work; all errors are $1\sigma$. While nominally our results favor a shallower power law, both values of $\alpha$ are consistent with one another within $1\sigma$ uncertainty. All are consistent with the $\alpha=-2.2$ expected from Kolmogorov turbulence.

As the timescales indicate that the stochastic variations are due to refractive scintillation, we can compute the modulation index from the time series of folded profile SNR:
\begin{equation}
m=\frac{\sigma_s}{\langle s\rangle}
\end{equation}
with $\sigma_s$ and $\langle s\rangle$ the standard deviation and mean of the profile SNR time series. We find $m\approx0.32$, somewhat lower than would be expected given previously measured modulation indices \citep{Rickett+Lyne1990} and the expected $m\propto f^{0.57}$ frequency scaling from Kolmogorov turbulence \citep{Lorimer+Kramer2012}. This computation is likely affected by the long-timescale trend seen over the course of the observing campaign and discussed in Section~\ref{sec:periodicity}.

%\begin{document}
\begin{table*}
  \renewcommand\thetable{1}
  \centering 
  \setlength{\tabcolsep}{1.5mm}
  \begin{tabular}{c c c}
  \hline
$f$ (MHz) & $\tau_{\mathrm{RISS}}$ (days) & Reference \\
\hline\hline
 73.8 & $>750$ & \quad\quad \cite{Rickett+Lyne1990} \\%[-0.3em] 
 111.5 & $>500$ & \quad\quad \cite{Rickett+Lyne1990} \\
 196 & $83\pm50$ & \quad\quad \cite{Rickett+Lyne1990} \\
 430 & $34\pm14$ & \quad\quad \cite{Rickett+Lyne1990} \\
 610 & $12\pm6$ & \quad\quad \cite{Rickett+Lyne1990} \\
 610 & $6.0\pm3.2$ & \quad\quad \cite{Rickett+Lyne1990} \\
 610 & $12$ & \quad\quad \cite{Stinebring+2000} \\
 1480 & $>0.54$ & \quad\quad \cite{Cordes+2004} \\
1550 & $2.85_{-0.60}^{+0.67}$ & \quad\quad This work (giant pulse rate) \\
 1550 & $3.27_{-1.51}^{+2.41}$ & \quad\quad This work (folded profile SNR) \\
 1668 & $0.162\pm0.004$ & \quad\quad \cite{Main+2021} \\
 2330 & $0.41\pm0.072$ & \quad\quad \cite{Cordes+2004} \\
\hline
  \end{tabular}
{\footnotesize
  \caption{Comparison with previous measurements of the Crab's refractive scintillation timescale. The timescales  derived from \citealt{Main+2021} and \citealt{Cordes+2004} were computed from the reported values of diffractive scintillation bandwidth and timescale, and error bars were computed in quadrature; all other timescales were computed via structure function analysis in their respective papers.\label{tab:previous-scintillation}}
  }
 
\end{table*} \label{tab:previous-scintillation}

\section{Periodicity} \label{sec:periodicity}

Previous studies (see e.g. \citealt{Mickaliger+2012}) reported on searches for periodicity in the Crab pulsar's giant pulse emission, albeit not with such a large selection of high-cadence data, like that from our observing campaign. \cite{Mickaliger+2012} found evidence for periodicities of 0.41 days and 0.99 days in observations at 330 MHz and 1.2 GHz, respectively, although the latter periodicity was found in randomized data and was therefore disregarded.

We performed a periodicity search on both the pulse rate and folded profile SNR time series, using the Lomb-Scargle analysis \citep{Lomb1976,Scargle1982} implemented in \texttt{astropy}'s\footnote{\url{https://www.astropy.org/}} \texttt{LombScargle} routine. We first removed all observations with a folded profile SNR below 6, leaving only observations with high-quality detections. We then computed periodograms from a minimum frequency of $f_{\mathrm{min}}=1/T$, with $T$ the length of the observing campaign, to $f_{\mathrm{max}}=2$ day$^{-1}$, and search for peaks. We chose to search the data at
\begin{equation}
N_{\mathrm{eval}}=n_oTf_{\mathrm{max}}
\end{equation}
frequencies, with $n_o=5$ \citep{VanderPlas2018}.

We then performed $10^5$ bootstrap simulations with each time series set. In each simulation, we kept all observing epochs the same, then randomly shuffle the pulse rates and profile SNRs. We used the results to compute power levels corresponding to $1\sigma$, $2\sigma$ and $3\sigma$ significance.

\begin{figure}
    \centering
    \includegraphics[width=0.95\columnwidth]{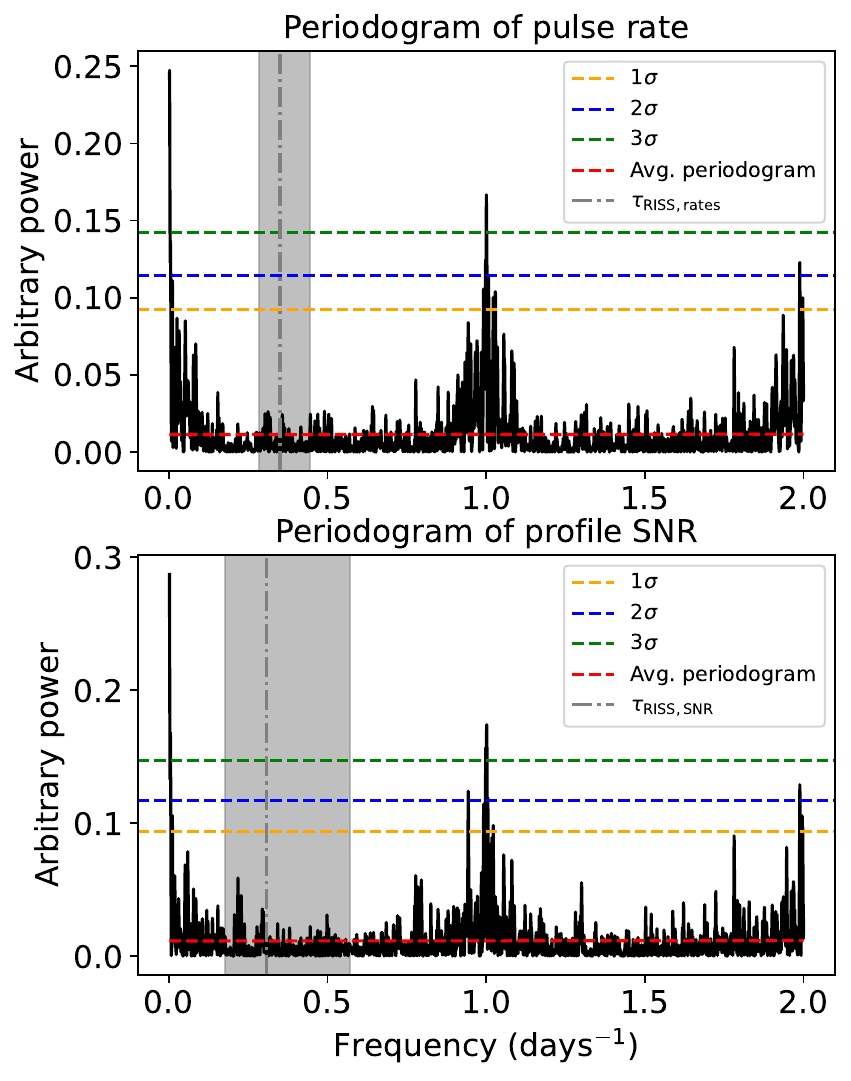}
    \caption{Periodograms for giant pulse rate and profile SNR. The orange, blue and green lines represent power significance levels of $1\sigma$, $2\sigma$ and $3\sigma$, respectively. The gray lines show the refractive timescales computed from each time series, with the gray shaded regions denoting the timescale uncertainties. The red lines shows the average periodogram of the $10^5$ simulations, and the gray line shows the refractive timescales for each time series. The average periodogram is actually not quite flat, with an increase towards higher frequencies, but the deviation from white noise is slight.}
    \label{fig:periodograms}
\end{figure}

Figure~\ref{fig:periodograms} displays the resulting periodograms. We see several peaks at the $3\sigma$ level. The lowest-frequency peaks correspond to $f_{\mathrm{min}}=1/T$ and frequencies slightly adjacent to it. The second set of peaks is close to a frequency of $1$ solar day$^{-1}$. This is interesting because it is close to the disregarded periodicity found at similar frequencies by \cite{Mickaliger+2012}. It is also close to the refractive timescale, resembling spurious peaks found by \cite{MacLeod+2010} in a study of variability of active galactic nuclei. The \cite{MacLeod+2010} peaks arose because of an underlying stochastic process coupled with the chosen observing cadence, and so it is tempting to attribute the peak to refractive scintillation. However, we see no peaks centered at frequencies corresponding to either of our derived values of the refractive timescale.

We suspect the peaks near $1$ solar day$^{-1}$ to be a consequence of two distinct effects: our observing schedule and aliasing. For the sake of consistency of optical depth, we attempted to observe the Crab pulsar at or near zenith throughout the campaign. In the summer months, the Crab lies close to the Sun, and was therefore observed during the hottest part of the day. In the winter, the Crab was observed during the coldest part of the night. This would have amplified temperature-dependent changes in system temperature with a timescale of one year, changing the measured SNR of both the folded profile and of individual pulses.

The degeneracy between the time of year and the time of day of our observations would have transformed this periodicity into an apparent periodicity of 1 solar day. The peak is further strengthened by aliasing \citep{VanderPlas2018}. With a period of 1 year and an observing cadence of 1 sidereal day, aliasing produces periodogram peaks at frequencies
\begin{equation}
f_{\mathrm{obs}}=f_{\mathrm{per}}+n\delta f
\end{equation}
with $f_{\mathrm{per}}=1$ yr$^{-1}$, $\delta f=1$ sidereal day$^{-1}$ and $n$ an integer.

Closer examination shows that the peaks at $\sim1$ day$^{-1}$ in both periodograms are composed of two subpeaks corresponding to the effects of observing at zenith and aliasing. To further support this, we give an example of how a simulated 1-year periodicity can result in an apparent 1-day periodicity in Appendix~\ref{sec:periodicity-appendix}.

% Effects of solar angle on observed giant pulse rates has been noted in other studies (e.g. \citealt{SerafinNadeau+2023}). The cluster of peaks near to $2$ day$^{-1}$ is then attributable to the next harmonic of this pattern.

% GD: The beam size of the 20m is large (~44 arcminutes for the L-band receiver), but we do avoid the Sun by, in almost all cases, at least 5 degrees. My hypothesis is that the periodicity is due to heating of the telescope itself, since the Sun doesn't pass close enough to the beam during the observing campaign.

We see no evidence for periodicities close to the $16.35$-day periodicity observed in the burst rate of FRB180916 \citep{CHIME2020} or the $\sim160$-day periodicity in burst rate proposed for FRB121102 \citep{Rajwade+2020,Cruces+2021}. Also of interest are the tentative $\mathcal{O}(1000\;\mathrm{day})$ periodicities found in pulse rates from the RRATs J1819--1458 and J1754--3014 \citep{Palliyaguru+2011}, but the total length of our observing campaign is too short to probe such long timescales.

We supplemented our Lomb-Scargle analyses with a Pearson chi-squared test (see e.g. \citealt{CHIME2020} for an example of its application). We folded each data set's pulses at a number of trial periods, assigning each pulse to one of $n$ phase bins. For each folding period, we computed the test statistic
\begin{equation}
\chi^2=\sum_{i=1}^{n}\frac{(O_i-E_i)^2}{E_i}
\end{equation}
where $O_i$ is the number of pulses in bin $i$ and $E_i=rT_i$, with $T_i$ the time length of bin $i$ and $r$ the mean pulse rate. We chose to fold at the periods corresponding to the frequencies searched in the Lomb-Scargle analysis. We again find evidence for the $1/T$ and $\sim1$ day$^{-1}$ apparent periodicities, along with higher harmonics.

\break

\section{Wait times} \label{sec:wait-times}

Our data set also allows us to test a long-standing assumption of Crab giant pulses. Giant pulses from the Crab pulsar are generally assumed to be a Poisson process; that is, the times between pulses are exponentially distributed \citep{Lundgren+1995,Karuppusamy+2010}:
\begin{equation}
    \mathcal{P}(\delta|r)=re^{-\delta r}
\end{equation}
where $\delta$ is the wait time between giant pulses, $r$ is the giant pulse rate, and $\mathcal{P}(\delta|r)$ is the conditional probability density for $\delta$ given some $r$.

However, the wait time distributions for some single-pulse sources, like FRB121102, are decidedly non-Poissonian. The wait times between bursts from FRB121102 is well-described by a Weibull distribution \citep{Opperman+2018}:
\begin{equation}
    \mathcal{W}(\delta|k,r)=k\delta^{-1}[\delta r\Gamma(1 + 1/k)]^ke^{-[\delta r\Gamma(1 + 1/k)]^k}
\end{equation}
where $k$ is a shape parameter and $\Gamma(x)$ is the gamma function defined by
\begin{equation}
    \Gamma(x)\equiv\int_0^\infty \mathrm{d}t\; t^{x-1}e^{-t}
\end{equation}
If $k=1$, the Weibull distribution reduces to the familiar exponential ditribution. A value $k<1$ indicates some sort of clustering in pulse arrival times; the other limit, $k\to\infty$, corresponds to a periodic time series. Some single-pulse sources besides FRB121102 exhibit clustering; for example, Vela giant micropulses appear to follow a Weibull distribution instead of an exponential distribution \citep{Chen+2020}.

Therefore, we searched for similar clustering in the arrival times of Crab giant pulses by assuming a Weibull distribution and allowing the possibility that $k\neq1$. We applied the method of \cite{Opperman+2018} to derive posterior distributions of $k$ and $r$ for each observation. For $N$ giant pulses arriving at times $t_1,...,t_N$, we computed a likelihood for the set of arrival times of
\begin{equation}
\begin{aligned}
    \mathcal{P}(N,t_1,...,t_N|k,r)=\;&r\;\mathrm{CDF}(t_1|k,r)\mathrm{CDF}(\Delta - t_N|k,r)\\
    &\times\prod_{i=1}^{N-1}\mathcal{W}(t_{i+1}-t_i|k,r)
\end{aligned}
\end{equation}
where $\Delta$ is the length of the observation and
\begin{equation}
    \mathrm{CDF}(\delta|k,r) \equiv e^{-[\delta r\Gamma(1 + 1/k]^k}
\end{equation}
is the cumulative distribution function of the Weibull distribution.\footnote{Under normal conventions, this quantity is actually $1$ minus the CDF; we follow the convention of \cite{Opperman+2018} for clarity.} We adopted Jeffreys priors for $k$ and $r$, i.e.
\begin{equation}
    \mathcal{P}(k,r)\propto k^{-1}r^{-1}
\end{equation}
and computed the joint posterior distribution by invoking Bayes' theorem:
\begin{equation}
    \mathcal{P}(k,r|N,t_1,...,t_N)\propto\mathcal{P}(N,t_1,...,t_N|k,r)\mathcal{P}(k,r)
\end{equation}
Individual posterior distributions for each of $k$ and $r$ can be obtained by marginalizing over the other quantity; error bars were computed by then integrating each individual posterior to find the interval encapsulating an area corresponding to a $1\sigma$ deviation.

\begin{figure}
    \centering
    \includegraphics[width=0.95\columnwidth]{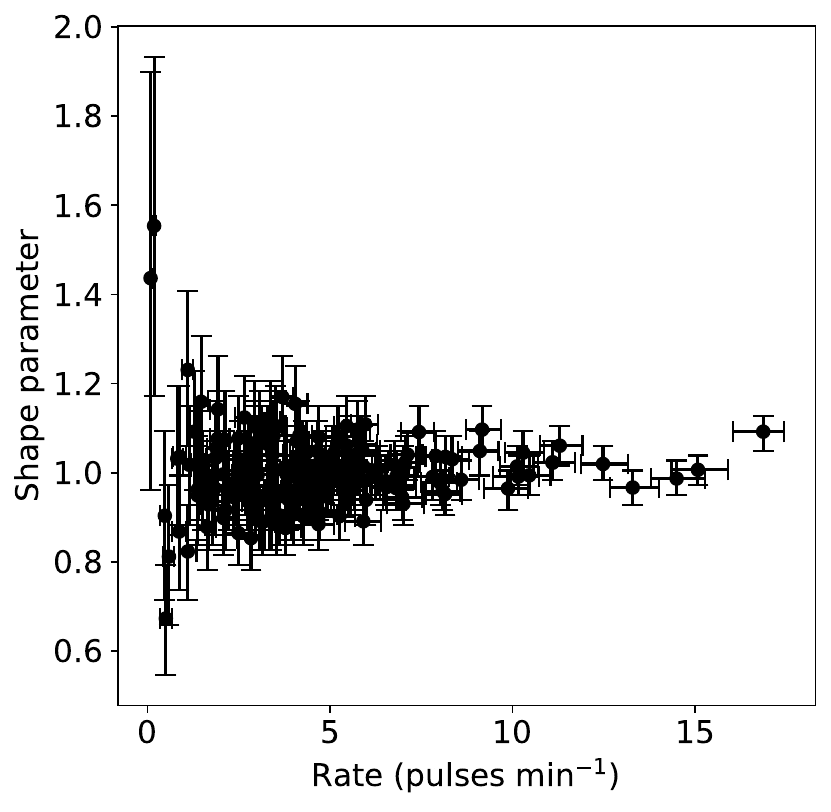}
    \caption{The Bayesian means of shape parameter $k$ and best-fit pulse rate $r$ computed for each observation. Error bars indicate $1\sigma$ uncertainties.}
    \label{fig:rate-shape-scatter}
\end{figure}

Figures~\ref{fig:rate-shape-scatter} illustrates the resulting best-fit pairs of $k$ and $r$ for each observation, determined by calculating the mean of each posterior. The majority of observations (122 of 176, or 69.3\%) are consistent with $k=1$ within $1\sigma$ errors, and 97.7\% are consistent within $2\sigma$, indicating little evidence of deviations from a Poisson process. The remaining four observations are likely contaminated by RFI; three show notably high pulse rates near 1675~MHz. All best-fit pulse rates arrived at by marginalizing over individual joint likelihoods are consistent with the actual measured pulse rate. Figure~\ref{fig:rate-shape-scatter} shows the mean values from each observation with $1\sigma$ error bars. We find no significant correlation between the Bayesian means of $k$ and $r$ (a Spearman coefficient of $\rho=0.079$).

\citet{Opperman+2018} note that in the limit of infinite spacing between observations, a likelihood for an observing campaign can be produced by multiplying together the likelihoods derived from each observation, treating them as independent. We do not do so here; the presence of refractive scintillation on timescales greater than the observational cadence means that successive observations are not independent.

\subsection{Pulse intensity and wait times} \label{subsec:intensity-wait-times}

\begin{figure}
    \centering
    \includegraphics[width=0.95\columnwidth]{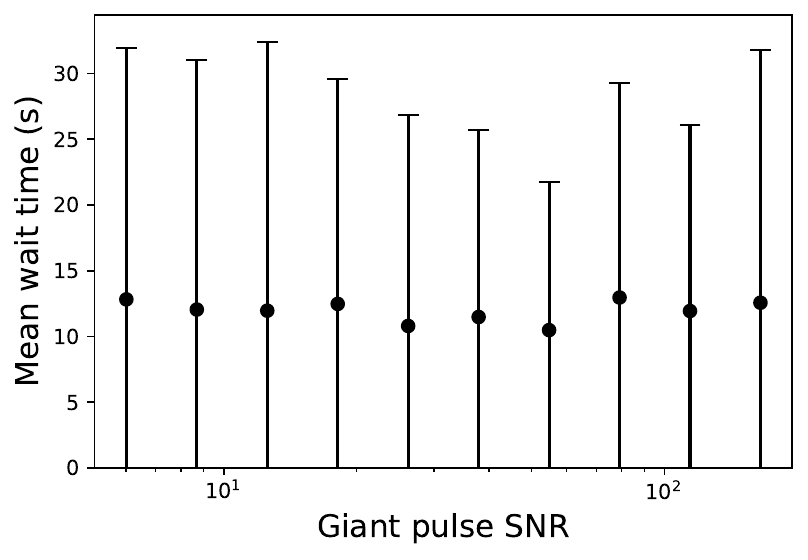}
    \caption{Wait times as a function of  giant pulse SNR, plotting on a logarithmic scale. Error bars represent the standard deviation of the wait times for the pulses in each SNR bin. The large standard deviations are due to the wide distribution of wait times.}
    \label{fig:snr-wait-times}
\end{figure}

Departures from a Poisson process could be due to relationships between a giant pulse's intensity and the wait time until the next pulse. We searched for this in both data sets. We binned the pulses logarithmically by its SNR, then took the mean of each bin $\bar{\delta}$. To quantify the relationship, we calculated both Pearson ($r$) and Spearman ($\rho$) correlation coefficients between $\log(\mathrm{SNR})$ and $\bar{\delta}$, and performed a least-square fit of a line to the trend. The results are shown in Figure~\ref{fig:snr-wait-times}. There is no significant correlation ($r=-0.09$, $\rho=0.17$) between giant pulse SNR and mean wait time.

\subsection{Diffractive scintillation} \label{subsec:autocorrelation}

Diffractive scintillation has the potential to affect observed pulse rates on short timescales. At 1668~MHz, \citealt{Main+2021} measured a diffractive timescale of $\Delta t_d=9.2\pm0.13$ seconds and a diffractive bandwidth of $\Delta\nu_{\mathrm{diff}}=1.10\pm0.02$~MHz, similar to earlier measurements \citep{Cordes+2004}. Since our bandwidth is much larger than the corresponding $\Delta\nu_{\mathrm{diff}}$ at 1550~MHz, our results should not be affected by diffractive scintillation.

\section{A possible event} \label{sec:event}

\begin{figure}
    \centering
    \includegraphics[width=0.95\columnwidth]{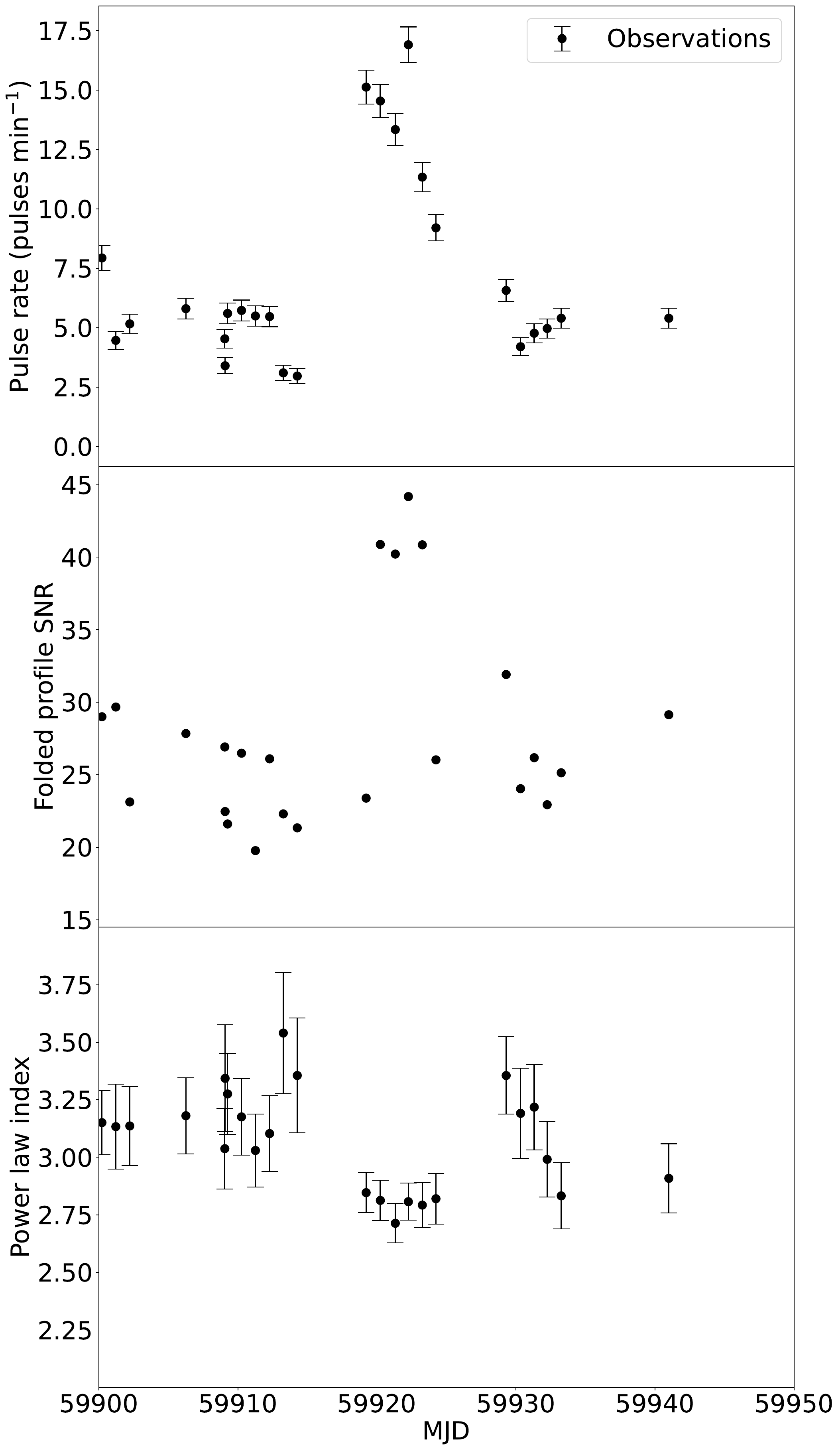}
    \caption{The giant pulse rate, and profile SNR and power-law index of giant pulse SNR distribution observed from the Crab shortly before, during, and after the event.}
    \label{fig:pulse-rate-anomaly}
\end{figure}

We detected  an interesting deviation in the Crab pulsar's activity in early December 2022 when its giant pulse rate increased to the highest level seen during the campaign, then regressed to a more typical value within two weeks via a roughly exponential decay, as shown in in Figure~\ref{fig:pulse-rate-anomaly}. This was accompanied by an increase in the strength of the folded profile, though without any clear similar trend. We also note that the power-law index of the giant pulse SNR distribution indicates that the distribution flattens during this period.

One possible explanation of the trend is refractive lensing similar to an extreme scattering event (ESE; see \citealt{Fiedler+1987}). ESEs have been observed in observations of both extragalactic radio sources and pulsars \citep{Cognard+1993,Coles+2015,Kerr+2018}, and refractive lensing has been invoked to explain a sharp increase and decrease in the observed burst rate of FRB20201124A \citep{Chen+2024}. Arising from inhomogeneities in the interstellar medium, ESEs often feature increases in the source's measured flux density bracketing a sharp dip of lengths extending to several years. Our observations end after the increase seen in the Crab's giant pulse rate and folded profile strength, meaning we do not have evidence for or against this sort of structure. We note that the possible flattening of the giant pulse SNR distribution is difficult to explain under this model.

The pattern also bears some resemblance to an event observed in 2018. The Crab pulsar is one of a number of pulsars known to exhibit glitches, characterized by sudden changes in  spin-down rate. So far, 30 glitches have been observed since the Crab pulsar's discovery.\footnote{See the glitch catalogues maintained by the Jodrell Bank Centre for Astrophysics (\url{https://www.jb.man.ac.uk/pulsar/glitches/gTable.html}) and the Australia Telescope National Facility (\url{https://www.atnf.csiro.au/people/pulsar/psrcat/glitchTbl.html}) for details.} The largest of these occurred on November 7, 2017, with a fractional spin frequency change of $\Delta\nu/\nu\approx5\times10^{-7}$ \citep{Shaw+2018}. \citet{Kazantsev+2019} reported the discovery of an increase in giant pulse rate and an even more extreme excess of strong giant pulses (fluence $>50000$ Jy ms) in a period beginning approximately five months after the glitch and lasting close to 100 days, raising the possibility that the glitch and the giant pulse excess were related. Evidence for an increase in pulse rate coupled to a glitch was also found for the RRAT J1819--1458, with a recovery in the absolute value of its frequency derivative, $|\dot{\nu}|$, similar to that seen after Crab glitches \citep{Lyne+2009}. No glitches from the Crab around this period have yet been reported, making this an unlikely explanation. Furthermore, this model does not explain the increase in profile SNR seen during the December 2022 event.

\section{Conclusions and future work} \label{sec:conclusion}

Studies of single pulses from neutron stars reveal interesting patterns of activity, including stochastic modulation and periodicity in pulse rates and/or fluxes. We present searches for some of these behaviors using a high-cadence observing campaign of the Crab pulsar, known for its giant pulse emission. We find a mean pulse of 4.73 pulses/minute, lower than most studies due to the observing frequency and the small collecting area of the 20m telescope. We find that these giant pulse rates show heavy stochastic modulation, and that the modulation timescales match extrapolations from low-frequency studies of refractive scintillation, indicating that the Crab is affected by Kolmogorov turbulence at GHz frequencies. We also show that the giant pulse rate can be used to probe this scintillation, a technique which can be applied to more sporadic single-pulse sources, such as fast radio bursts, and find that two methods of compensating for it produce similar results.

We find no evidence for periodicity in either giant-pulse rate or SNR, as seen in certain FRBs and RRATs, besides excess power at periods of roughly 1~day$^{-1}$ that can likely be attributed to systematic effects. We also find that within individual observations, giant pulses do not show evidence of clustering in time, and that there is at most a mild correlation between pulse strength and pulse-to-pulse wait times, with Pearson and Spearman correlation coefficients of $r=-0.09$ and $\rho=0.17$, respectively.

We also find a possible change in giant pulse rate in early December 2022, followed by a gradual return to normal levels over the course of $\sim2$ weeks. The pattern is similar to that observed in the rate of strong giant pulses several months after the 2017 glitch, although in the absence of reports of any other sudden changes in the Crab pulsar's behavior around the time of the 2022 event, we have no evidence that it may be connected to a glitch or other rotational anomaly. An alternative explanation is that this is the beginning of a refracting lensing event similar to ESEs observed in observations of other pulsars, although our observations do not cover the date range which would show the expected subsequent evolution in flux density.

We are currently performing a second daily cadence observing campaign targeting the Crab to address two limitations of the first. In addition to extending the number of giant pulses to make our statistical results more robust, the extended baseline will allow us to determine whether the apparent periodicities found in Section~\ref{sec:periodicity} are indeed seasonal artefacts. We are also performing pairs of observations on certain days, allowing us to compute the structure function at low time lags and reach the noise-dominated regime, putting better constraints on the refractive timescale.

\subsection*{Software}

This work made use of several software packages, including \presto\;\citep{Ransom2011}, \texttt{DSPSR}\;\citep{DSPSR}, \texttt{PSRCHIVE}\;\citep{PSRCHIVE} and \texttt{astropy} \citep{Astropy}. This work has also made use of NASA's Astrophysics Data System.

% GD: The above phrasing is what ADS requests; c.f. https://ui.adsabs.harvard.edu/about/.

\acknowledgments

GMD was supported through NSF award \# 2108673. GMD, NL, AS, CS and MAM are members of the NANOGrav PFC, supported by NSF award \# 2020265. The authors also wish to thank the Pulsar Science Collaboratory for enabling the observations, and the anonymous referee for their helpful comments.

% GD: That last sentence is *terrible* writing, sorry! I will revise it but want to make sure I don't forget to add something there.

\appendix

\section{Periodicity simulations} \label{sec:periodicity-appendix}

To demonstrate how the 1-day periodicity may be due to both observing at zenith and yearly system temperature variations, we consider a simulated signal of the form
\begin{equation}
X(t) = A\sin(\omega t) + c + \mathrm{white\;noise}
\end{equation}
where $A=1$ pulses min$^{-1}$, $\omega=2\pi/(1\;\mathrm{year})$, and $c=4.73$ pulses min$^{-1}$, and the white noise has mean $0$ and standard deviation $1$ pulses min$^{-1}$. The parameters are chosen such that the mean and standard deviation of the time series are similar to those of the actual time series of giant pulse rate, although our results below still hold if we vary these parameters.

\begin{figure}[h]
    \centering
    \includegraphics[width=0.75\columnwidth]{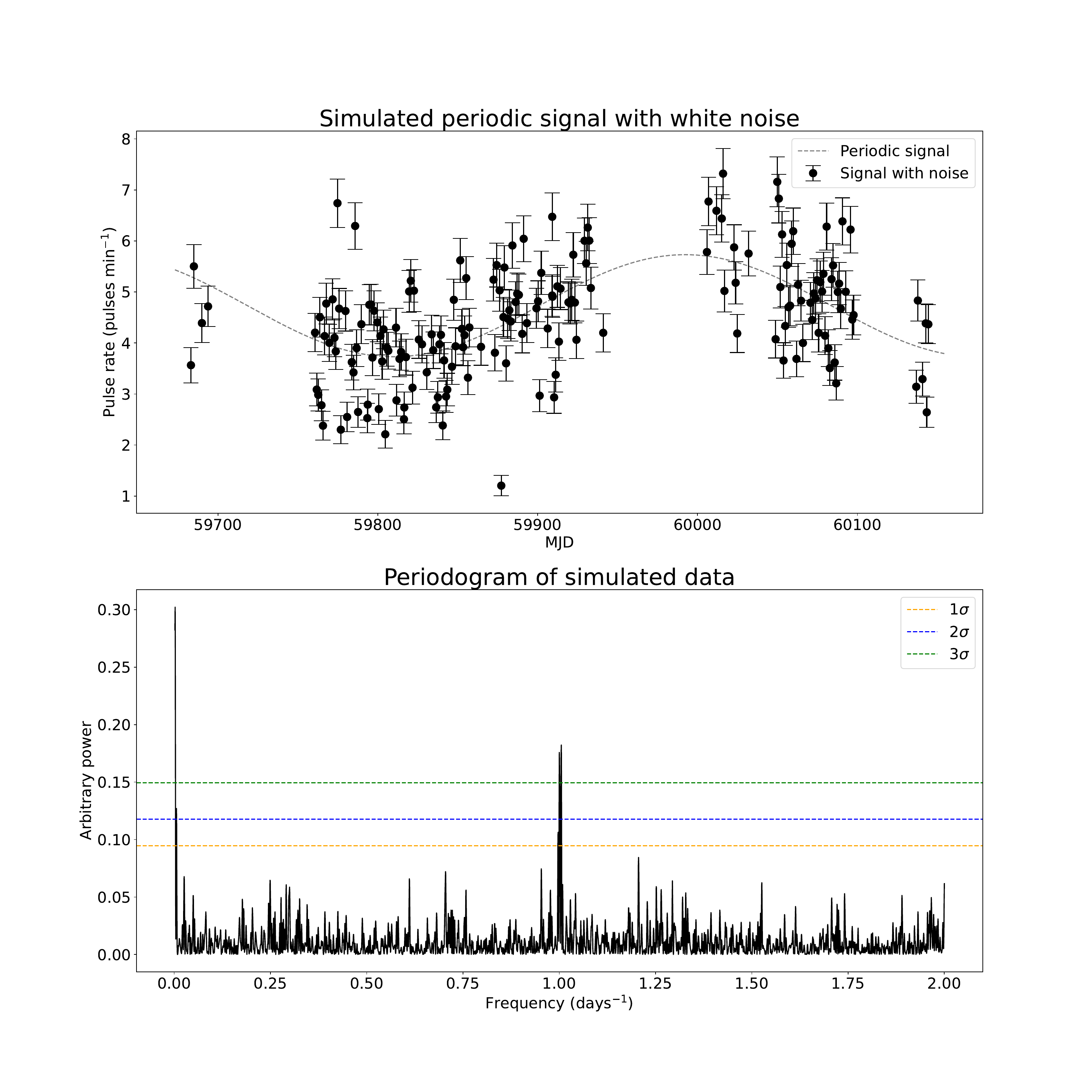}
    \caption{A simulated sinusoidal signal with properties similar to the observed giant pulse rate from the Crab, and the Lomb-Scargle periodogram of the time series.}
    \label{fig:simulated-data}
\end{figure}

\begin{figure}[h]
    \centering
    \includegraphics[width=0.75\columnwidth]{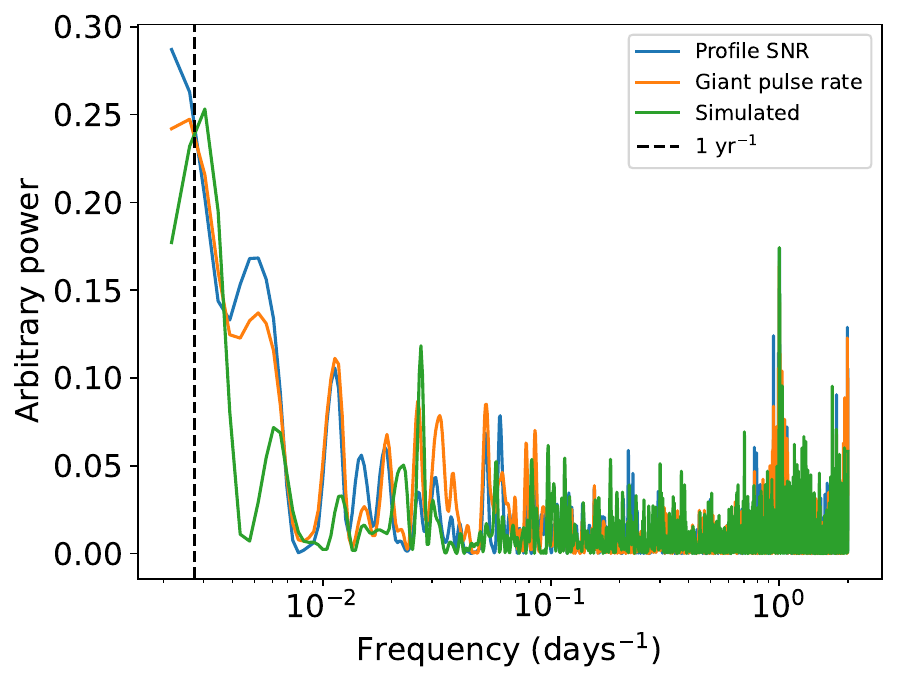}
    \caption{A logarithmic plot of the two periodograms from the real data and a periodogram from the simulated data. The bin at $1$ yr$^{-1}$ has either a peak of its own or an excess distinct from the power contained at the bin at $1/T$.}
    \label{fig:periodogram-comparison}
\end{figure}

Figure~\ref{fig:simulated-data} displays a time series of this simulated data, with mock observations taken at the same times as our actual observations. The accompanying periodogram closely resembles the periodograms of the real signal. We again see peaks at $1$ year$^{-1}$ and higher harmonics, as well as near $1$ day$^{-1}$ at the two frequencies expected from zenith observations and from aliasing. Figure~\ref{fig:periodogram-comparison} makes the peak at $1$ yr$^{-1}$ clearer and supports the aliasing explanation.

\raggedbottom

\clearpage

\section{Observation details} \label{sec:observation-appendix}

Table~\ref{tab:observation-details} includes the information required to reproduce most of the time series analysis conducted in this paper. Additional information is available upon request. Observation MJDs are listed to two decimals places to differentiate between multiple observations conducted close together on the same day.

\begin{longtable*}{cccc}
\caption{Details of each observation from the campaign.}\label{tab:observation-details}\\\hline
Date of observation & Profile SNR & $N_{\mathrm{GP}}$ & Giant pulse rate\\
MJD & & & Pulses min$^{-1}$\\\hline
\endhead
59683.04 & 8.56 & 92 & 3.07 \\
59684.88 & 19.31 & 126 & 4.20 \\
59689.90 & 27.32 & 175 & 5.83 \\
59693.74 & 19.03 & 104 & 3.47 \\
59760.69 & 12.65 & 28 & 0.93 \\
59761.75 & 17.79 & 72 & 2.40 \\
59762.69 & 15.25 & 167 & 5.57 \\
59763.71 & 9.07 & 35 & 1.17 \\
59764.71 & 24.94 & 144 & 4.80 \\
59765.65 & 19.23 & 134 & 4.46 \\
59766.68 & 14.56 & 55 & 1.83 \\
59767.70 & 27.23 & 142 & 4.73 \\
59769.66 & 12.24 & 36 & 1.20 \\
59771.70 & 28.19 & 198 & 6.60 \\
59772.84 & 19.58 & 65 & 2.17 \\
59773.65 & 19.61 & 104 & 3.46 \\
59774.72 & 16.00 & 85 & 2.83 \\
59775.77 & 15.00 & 57 & 1.90 \\
59776.80 & 11.52 & 28 & 0.93 \\
59779.75 & 6.65 & 6 & 0.20 \\
59780.75 & 9.02 & 16 & 0.53 \\
59783.69 & 9.46 & 20 & 0.67 \\
59784.74 & 12.56 & 43 & 1.43 \\
59785.77 & 17.66 & 58 & 1.93 \\
59786.72 & 17.96 & 40 & 1.33 \\
59787.61 & 16.76 & 72 & 2.40 \\
59789.67 & 16.92 & 88 & 2.93 \\
59793.39 & 9.98 & 34 & 1.13 \\
59793.66 & 12.02 & 45 & 1.50 \\
59794.64 & 16.04 & 43 & 1.43 \\
59795.76 & 17.50 & 51 & 1.70 \\
59796.59 & 23.94 & 109 & 3.63 \\
59797.60 & 9.73 & 17 & 0.57 \\
59799.62 & 18.58 & 52 & 1.73 \\
59800.56 & 19.29 & 124 & 4.13 \\
59801.58 & 17.06 & 64 & 2.13 \\
59802.70 & 11.95 & 64 & 2.13 \\
59803.55 & 28.28 & 246 & 8.20 \\
59804.63 & 24.52 & 213 & 7.10 \\
59805.60 & 29.49 & 180 & 6.00 \\
59806.54 & 30.30 & 236 & 7.86 \\
59811.35 & 11.43 & 43 & 1.43 \\
59811.64 & 24.53 & 148 & 4.93 \\
59813.53 & 18.33 & 108 & 3.60 \\
59814.53 & 28.52 & 191 & 6.36 \\
59816.31 & 21.55 & 165 & 5.49 \\
59816.53 & 25.49 & 172 & 5.73 \\
59817.57 & 19.76 & 104 & 3.47 \\
59819.58 & 21.38 & 43 & 1.43 \\
59820.61 & 17.81 & 86 & 2.86 \\
59821.68 & 23.46 & 124 & 4.13 \\
59822.57 & 31.07 & 168 & 5.60 \\
59825.62 & 27.32 & 215 & 7.17 \\
59827.60 & 23.65 & 135 & 4.50 \\
59830.58 & 20.05 & 87 & 2.90 \\
59833.65 & 15.76 & 76 & 2.53 \\
59834.63 & 18.26 & 107 & 3.57 \\
59836.47 & 19.35 & 57 & 1.90 \\
59837.53 & 15.92 & 58 & 1.93 \\
59838.51 & 19.27 & 74 & 2.47 \\
59839.47 & 15.41 & 68 & 2.26 \\
59840.59 & 21.14 & 141 & 4.70 \\
59841.43 & 17.49 & 83 & 2.77 \\
59842.66 & 15.21 & 100 & 3.33 \\
59843.43 & 27.63 & 143 & 4.76 \\
59846.38 & 18.90 & 110 & 3.67 \\
59847.51 & 20.47 & 162 & 5.39 \\
59848.62 & 19.58 & 112 & 3.73 \\
59851.52 & 10.71 & 26 & 0.87 \\
59852.58 & 19.43 & 96 & 3.20 \\
59853.32 & 21.75 & 105 & 3.50 \\
59854.21 & 24.50 & 142 & 4.73 \\
59855.27 & 20.85 & 130 & 4.33 \\
59856.35 & 26.40 & 120 & 4.00 \\
59857.26 & 25.77 & 140 & 4.67 \\
59864.54 & 2.78 & 116 & 3.86 \\
59872.36 & 23.94 & 154 & 5.13 \\
59873.23 & 29.72 & 158 & 5.26 \\
59874.30 & 29.95 & 139 & 4.63 \\
59876.18 & 24.39 & 90 & 3.00 \\
59877.20 & 18.55 & 93 & 3.10 \\
59878.56 & 27.62 & 74 & 2.46 \\
59879.18 & 21.98 & 81 & 2.70 \\
59880.21 & 20.54 & 141 & 4.70 \\
59881.25 & 23.11 & 94 & 3.13 \\
59882.26 & 19.04 & 130 & 4.33 \\
59883.18 & 28.70 & 176 & 5.87 \\
59884.18 & 26.74 & 310 & 10.32 \\
59886.20 & 36.57 & 334 & 11.14 \\
59887.18 & 40.57 & 306 & 10.20 \\
59888.22 & 39.49 & 315 & 10.50 \\
59890.33 & 27.16 & 305 & 10.17 \\
59891.22 & 31.38 & 212 & 7.06 \\
59893.25 & 29.39 & 204 & 6.80 \\
59899.23 & 23.18 & 224 & 7.47 \\
59900.22 & 29.00 & 238 & 7.94 \\
59901.22 & 29.67 & 134 & 4.47 \\
59902.23 & 23.12 & 155 & 5.16 \\
59906.26 & 27.84 & 174 & 5.80 \\
59909.06 & 26.92 & 136 & 4.53 \\
59909.08 & 22.47 & 102 & 3.40 \\
59909.26 & 21.61 & 168 & 5.60 \\
59910.26 & 26.49 & 172 & 5.73 \\
59911.26 & 19.77 & 165 & 5.49 \\
59912.28 & 26.10 & 164 & 5.47 \\
59913.27 & 22.30 & 93 & 3.10 \\
59914.27 & 21.34 & 89 & 2.97 \\
59919.23 & 23.39 & 453 & 15.12 \\
59920.25 & 40.87 & 436 & 14.54 \\
59921.32 & 40.22 & 400 & 13.34 \\
59922.26 & 44.18 & 507 & 16.90 \\
59923.26 & 40.85 & 340 & 11.34 \\
59924.25 & 26.03 & 276 & 9.20 \\
59929.30 & 31.91 & 197 & 6.57 \\
59930.33 & 24.04 & 126 & 4.20 \\
59931.31 & 26.17 & 143 & 4.77 \\
59932.24 & 22.93 & 149 & 4.97 \\
59933.25 & 25.13 & 162 & 5.40 \\
59940.99 & 29.14 & 162 & 5.40 \\
60005.88 & 26.14 & 206 & 6.86 \\
60006.89 & 21.72 & 89 & 2.97 \\
60011.85 & 21.53 & 211 & 7.03 \\
60015.17 & 24.83 & 159 & 5.30 \\
60015.97 & 2.74 & 275 & 9.15 \\
60016.83 & 27.96 & 246 & 8.20 \\
60022.88 & 33.22 & 335 & 11.14 \\
60023.86 & 26.86 & 3 & 0.10 \\
60024.84 & 37.91 & 298 & 9.92 \\
60031.94 & 14.51 & 59 & 1.97 \\
60048.85 & 23.15 & 179 & 5.97 \\
60049.88 & 24.01 & 179 & 5.97 \\
60050.87 & 24.25 & 162 & 5.39 \\
60051.81 & 29.30 & 117 & 3.90 \\
60052.92 & 20.02 & 141 & 4.70 \\
60053.81 & 23.05 & 123 & 4.09 \\
60054.88 & 17.69 & 76 & 2.53 \\
60055.84 & 17.25 & 66 & 2.20 \\
60056.91 & 24.11 & 105 & 3.50 \\
60057.91 & 21.25 & 126 & 4.20 \\
60058.89 & 23.30 & 124 & 4.13 \\
60059.86 & 18.86 & 82 & 2.73 \\
60061.90 & 28.74 & 252 & 8.38 \\
60062.93 & 26.55 & 189 & 6.30 \\
60064.77 & 21.86 & 114 & 3.79 \\
60065.95 & 18.71 & 119 & 3.96 \\
60070.66 & 20.56 & 90 & 2.99 \\
60071.82 & 21.09 & 115 & 3.83 \\
60072.75 & 23.83 & 169 & 5.63 \\
60073.79 & 20.34 & 163 & 5.43 \\
60074.74 & 31.08 & 242 & 8.07 \\
60075.77 & 32.94 & 376 & 12.52 \\
60076.78 & 36.36 & 197 & 6.57 \\
60077.92 & 20.62 & 178 & 5.93 \\
60078.78 & 17.16 & 132 & 4.40 \\
60079.81 & 23.79 & 103 & 3.43 \\
60080.78 & 36.25 & 259 & 8.64 \\
60081.82 & 24.57 & 244 & 8.14 \\
60082.78 & 23.17 & 212 & 7.06 \\
60083.85 & 24.70 & 137 & 4.56 \\
60084.80 & 21.72 & 102 & 3.40 \\
60085.82 & 26.58 & 181 & 6.03 \\
60086.77 & 21.84 & 125 & 4.17 \\
60087.78 & 25.53 & 170 & 5.67 \\
60088.57 & 30.64 & 206 & 6.87 \\
60089.65 & 27.88 & 143 & 4.76 \\
60090.69 & 23.55 & 106 & 3.53 \\
60092.73 & 27.67 & 142 & 4.73 \\
60095.74 & 15.07 & 47 & 1.57 \\
60096.81 & 22.02 & 122 & 4.07 \\
60097.64 & 26.47 & 102 & 3.40 \\
60136.84 & 13.86 & 59 & 1.96 \\
60137.85 & 24.21 & 95 & 3.17 \\
60140.76 & 22.82 & 129 & 4.30 \\
60142.80 & 17.44 & 116 & 3.86 \\
60143.42 & 20.73 & 77 & 2.57 \\
60144.42 & 23.29 & 107 & 3.57 \\
\hline
Total & 5281.68 & 24985 & 4.73 \\
%\label{tab:PSC-multipage-obs}
\end{longtable*}

\break

\bibliography{giant_pulses.bib}{}
\bibliographystyle{aasjournal}

\end{document}